\documentclass[doublecol]{epl2} 

\usepackage{graphicx,epsf,bm,bbm,color,amsmath,amssymb,ulem}

	\newcommand{\beq}{\begin{equation}}
	\newcommand{\be}{\begin{equation}}
	\newcommand{\beqn}{\begin{eqnarray}}
	\newcommand{\eeq}{\end{equation}}
	\newcommand{\ee}{\end{equation}}
	\newcommand{\eeqn}{\end{eqnarray}}

\title{Quantum loss of synchronization in the dynamics of two spins}

\author{Y. Liu\inst{1} \and F. Pi\'echon\inst{1} \and J.N. Fuchs\inst{2,1}}
\shortauthor{Y. Liu \etal}

\institute{                    
  \inst{1} Laboratoire de Physique des Solides, CNRS UMR 8502, Univ. Paris-Sud, F-91405 Orsay\\
  \inst{2} Laboratoire de Physique Th\'eorique de la Mati\`ere Condens\'ee, CNRS UMR 7600, Universit\'e Pierre et Marie Curie, 4 place Jussieu, F-75252 Paris
}

\abstract{
Motivated by the spin self-rephasing recently observed in an atomic clock, 
we introduce a simple dynamical model to study the competition between dephasing and synchronization. 
Two spins $S$ are taken to be initially parallel and in the plane perpendicular to an inhomogeneous 
magnetic field $\Delta$ that tends to dephase them. In addition, the spins are coupled by exchange interaction $J$ 
that tries to keep them locked. The analytical solution of the classical dynamics shows that, 
there is a phase transition to a synchronized regime for sufficiently large exchange interaction $J>\Delta$ compared to the inhomogeneity. The quantum dynamics is solved analytically in four limits -- large/small $J/\Delta$ and large/small $S$ -- and numerically in between. In sharp contrast to the classical case, the quantum solution features very rich $S$-dependent multiscale dynamics. For any finite $S$, there is no synchronization but a crossover around $J=\Delta$ between two regimes. The synchronization transition is only recovered when $S\to \infty$, approaching the classical solution in a non-trivial way. Quantum effects therefore suppress the synchronization transition.}

\begin{document}

\maketitle

\section{Introduction}
Synchronization is a collective phenomenum that occurs in the dynamics of many different systems, see e.g. \cite{pedersen} 
and references therein. A famous classical model of synchronization is that introduced by Kuramoto
 \cite{kuramoto,originalkuramoto}. In this model, two or more oscillators with distinct frequencies can synchronize 
when they are coupled by a sufficiently strong non-linearity.  
Recently, a synchronization transition was observed in a quantum systems made of a large ensemble of spins $1/2$ 
\cite{deutsch,kleine}. In an atomic clock, $N\sim 4.10^4$ trapped two-level atoms behaving as pseudo-spins were found to synchronize beyond 
a critical density or interaction strength. In the experiment, the contrast of the Ramsey fringes measures the coherence of the atomic clock 
and typically decays in time. It was found that this decay almost stops at sufficiently large density, substantially increasing the coherence of the clock. 
A simple picture explaining this self-rephasing was proposed in \cite{deutsch}.
It involves two equal populations of atoms -- corresponding to hot and cold atoms -- each represented by a macrospin $S=N/4$. 
The two macrospins feel different longitudinal magnetic fields, because of the spatial inhomogeneity of the atomic cloud. 
In addition, atom-atom collisions generate an effective exchange coupling for the two macrospins. When the latter is strong enough, it impedes dephasing of the macrospins.

Building further on this picture, we introduce a quantum model of two (macro-)spins $S$ to describe the dynamical competition between dephasing 
by the inhomogeneous magnetic field and synchronization by exchange interaction. A legitimate question to ask is whether 
a genuine synchronization transition exists in a {\it quantum} model \cite{gibble}. Here, we answer that question by presenting our results, 
leaving details of derivations to a longer companion paper \cite{long}. We start by introducing the model, before giving its classical solution 
and several approximate quantum solutions. Our aim is to provide a qualitative picture of the quantum dynamics as a function of the spin size $S$ 
and of the ratio between exchange and inhomogeneity. 

\section{Two spins model}
We consider the dynamics of two spins $\vec{S}_{1},\vec{S}_{2}$, of size $S$, 
coupled by exchange interaction and subjected to an inhomogeneous magnetic field in the $z$ direction.
The corresponding Hamiltonian ressembles that of the two-level BCS model \cite{dusuel}
and reads $H=J_s\vec{S}_1\cdot \vec{S}_2+ \Delta_s(S_1^z-S_2^z)$ 
with $J_s$ (resp. $\Delta_s$) the characteristic exchange (resp. inhomogeneity) energy (hereafter $\hbar=1$).
The initial state is taken to be a coherent state in the direction $x$ perpendicular 
to the magnetic field: $|\psi(0)\rangle=|S_1^x=S,S_2^x=S\rangle$.
We characterize the state $|\psi(t)\rangle$ at time $t$ by three  quantities:  the single spin contrast $C_1(t)=\frac{|\langle \vec{S}_{1,2} \rangle|}{S}$, 
the single spin unit vector direction $\vec{n}_{1,2}(t)=\frac{\langle \vec{S}_{1,2} \rangle}{|\langle \vec{S}_{1,2} \rangle|}$ and the total spin contrast 
$C(t)=\frac{|\langle \vec{S}_{1} +\vec{S}_{2}\rangle|}{2S}$, the latter corresponding to the quantity that is measured 
in Ramsey fringes experiments \cite{deutsch,kleine}. 
At any time we can write
$C(t)=C_1(t)\sqrt{(1+\vec{n}_1\cdot \vec{n}_2)/2}$.
For classical spins $C_1(t)=1$ since $\vec{S}_{1,2}(t)=S\vec{n}_{1,2}(t)$.
For quantum spins  $C_1(t)$ is in fact a direct measure of the effective spreading width  $D_1(t)=\frac{\langle(\vec{S}_1-\langle \vec{S}_1\rangle)^2\rangle}{S^2}$
of the single spin wavepackets; indeed, quite generally $C_1(t)=1-(D_1(t)-\frac{1}{S})$.
For our specific choice of initial state, the initial values are  
 $C(0)=C_1(0)=1$, $D_1(0)=\frac{1}{S}$ and $\vec{n}_{1,2}(0)=\vec{e}_x$; 
furthermore at any time we also have $\vec{n}_2(t)=[n_1^x,-n_1^y,-n_1^z]$ such that we can rewrite $C(t)=C_1(t)|n_1^x(t)|$.

\section{Classical synchronization transition}
\begin{figure}[ht]
\onefigure[width=6cm]{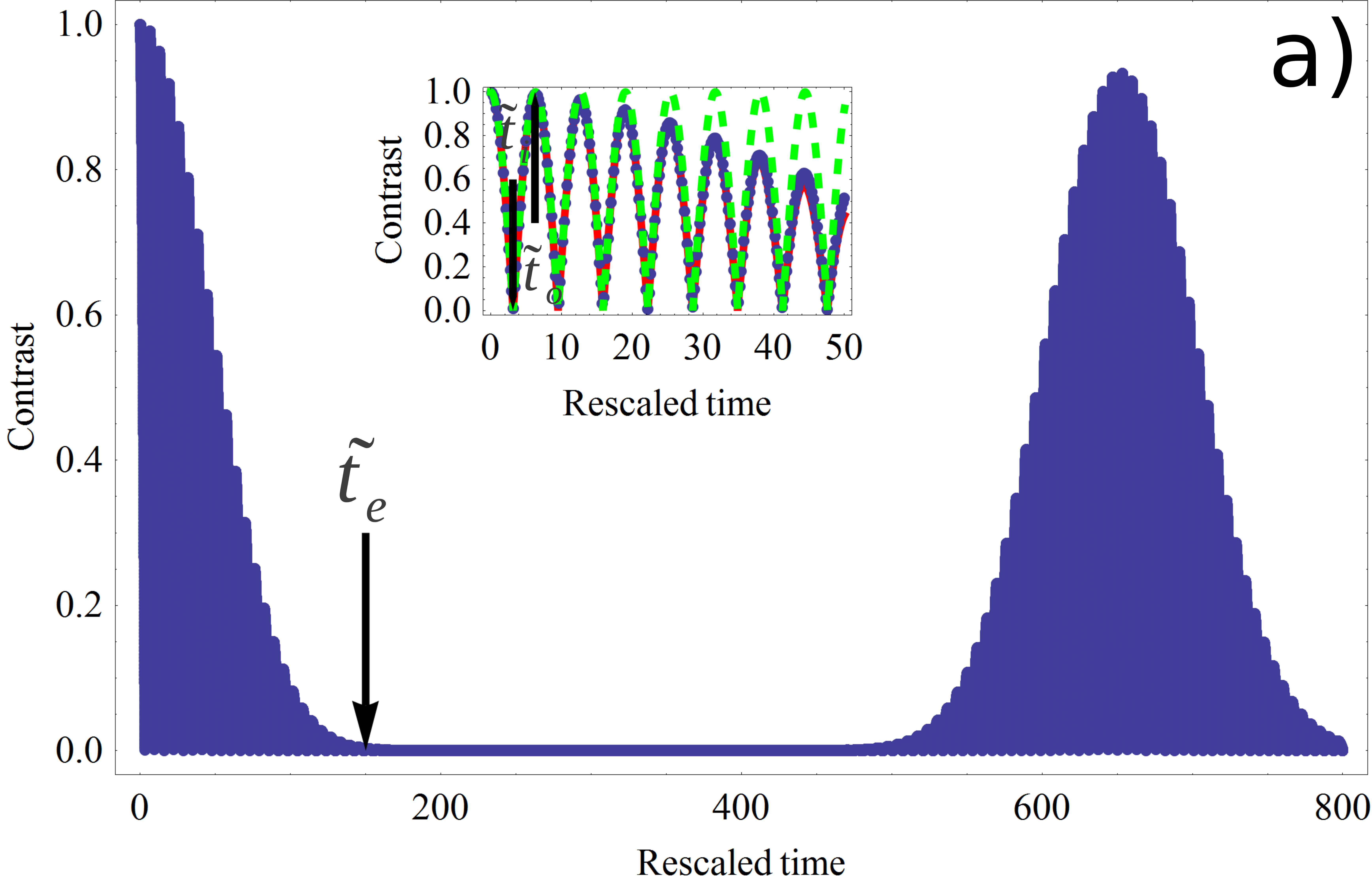}
\onefigure[width=6cm]{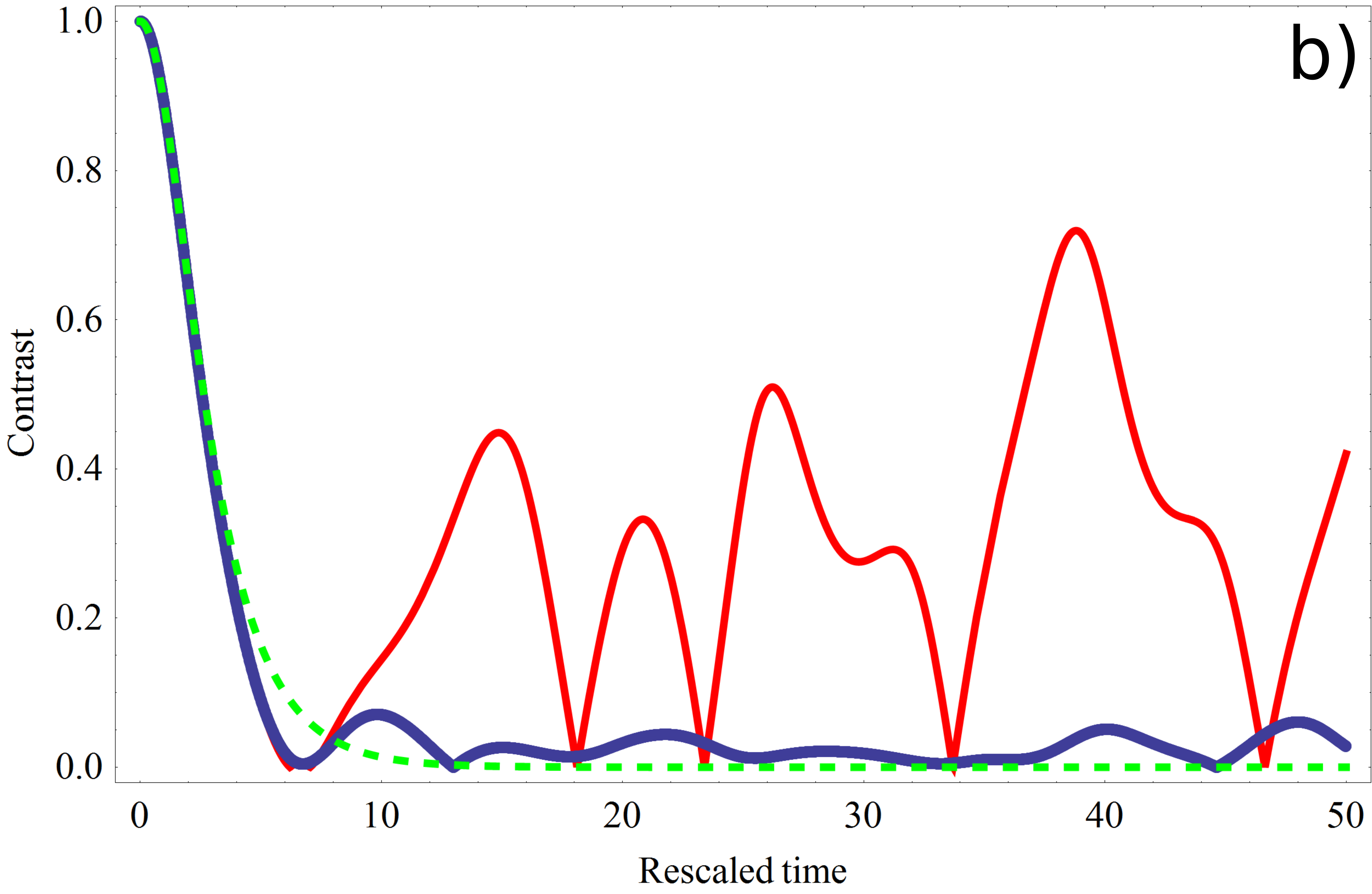}
\onefigure[width=6cm]{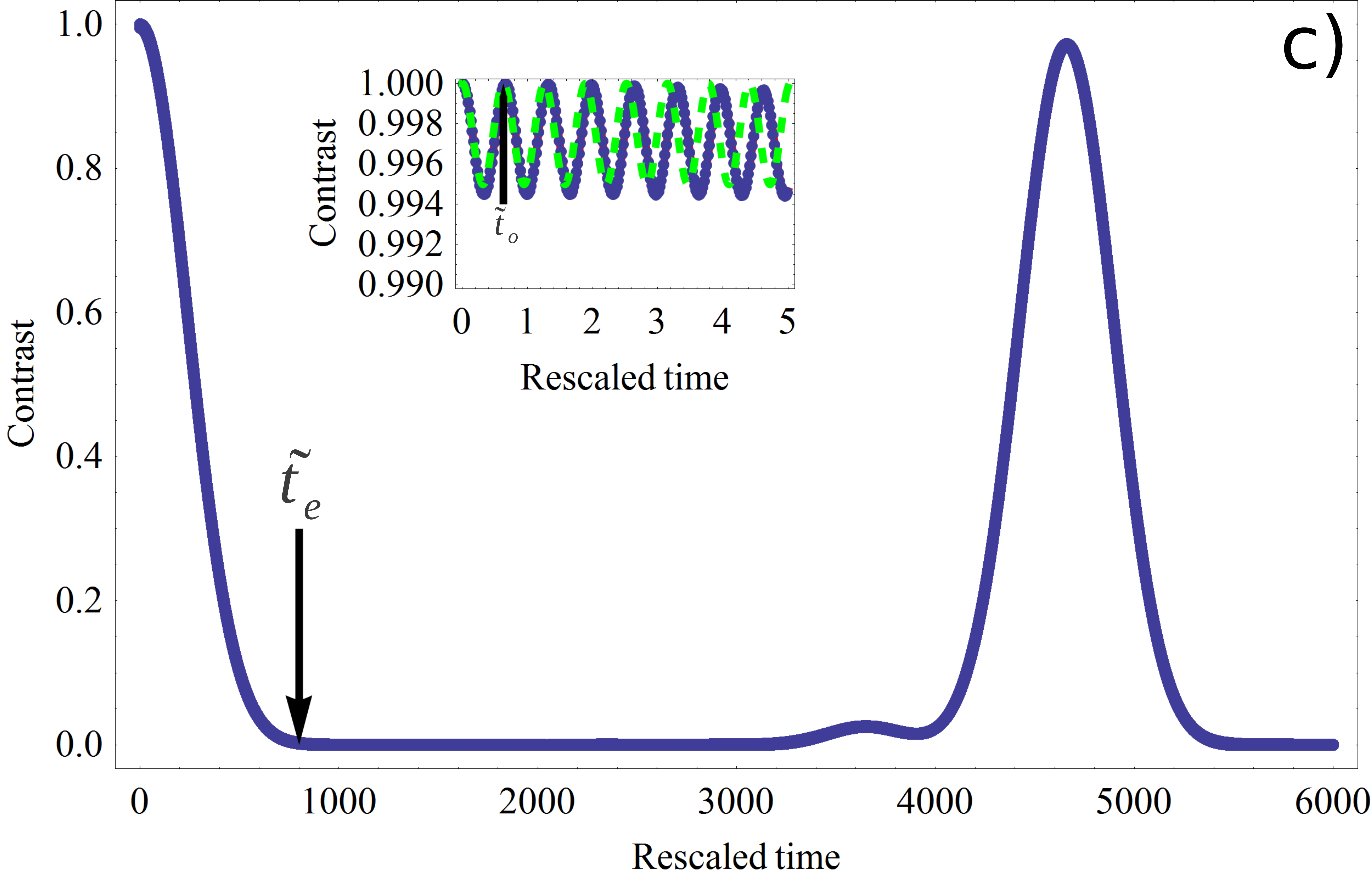}
 \caption{Contrast $C$ as a function of rescaled time $\tilde{t}$ [units of $\frac{2}{\Delta}$] for $S=10$. 
Comparison between classical dynamics (dashed green line), quantum dynamics (dotted blue line) and cumulant 
approach (full red line) in 3 regimes: (a) $J/\Delta=0.2$ (dephased); (b) $J/\Delta=1$ (critical); (c) $J/\Delta=10$ (synchronized). 
Initial decay ($\tilde{t}_i$), fast oscillations ($\tilde{t}_o$) and envelope ($\tilde{t}_e$) times are indicated.}
\label{fig1}
\end{figure}
We first consider the classical version of the two spins model. We get rid of the spin size $S$ by defining rescaled quantities: time 
$\tilde{t}\equiv \Lambda_s^{-1}t$, exchange $J\equiv \Lambda_s ^2 J_s $, inhomogeneity $\Delta\equiv \Lambda_s \Delta_s$ and spins $\vec{n}_{1,2}\equiv \Lambda_s^{-1}\vec{S}_{1,2}$ 
with $\Lambda_s=S$.
Writing $\vec{n}_{1}=[\cos \theta,\sin \theta \sqrt{1-(\frac{J}{\Delta})^2\sin^2\theta},\frac{J}{\Delta}\sin^2 \theta]$, 
the spin equations of motion $\dot{\vec{n}}_{1}=[\Delta \vec{e}_z +J\vec{n}_{2}(\tilde{t})]\times \vec{n}_{1}(\tilde{t})$ 
map to a nonlinear pendulum equation $\dot{\theta}(\tilde{t})=  \Delta\sqrt{1-\frac{J^2}{\Delta^2} \sin^2 \theta(\tilde{t})}$ \cite{long}.
The solution $\theta(\tilde{t})$ is obtained as the incomplete elliptic integral of the first kind $F(\sin{\theta};\frac{J}{\Delta}) = \Delta\tilde{t} $ such that the contrast
is 
$C(\tilde t)=\left|\textrm{cn}[\Delta\tilde{t};\frac{J}{\Delta}]\right|$
with $\textrm{cn}$ the Jacobi elliptic function.
At short time, $\theta(\tilde t)\simeq \Delta\tilde{t}$ is linear in time
such that $C(\tilde{t})\simeq 1-\frac{\theta^2}{2}\approx 1-\tilde{t}^2/\tilde{t}_i^2$ with $\tilde{t}_{i}=\frac{\sqrt{2}}{\Delta}$.
For larger time the dynamics is nonlinear and, depending on the ratio $\frac{J}{\Delta}$, we distinguish three regimes. The case $J<\Delta$ corresponds to the dephased regime; the contrast vanishes periodically with a period 
$\tilde{t}_p=\frac{2}{\Delta}K(\frac{J}{\Delta})\approx \frac{\pi}{\Delta} $  with $K(k)$  
the complete elliptic integral of the first kind (Fig. \ref{fig1}a, green dashed). 
The case $J>\Delta$ corresponds to  the synchronized regime, which, by definition, means that the contrast remains finite at all times. It reaches its minimal value $C_{\textrm{min}}=\sqrt{1-\frac{\Delta^2}{J^2}}$ periodically with a period
$\tilde{t}_p=\frac{2}{J}K(\frac{\Delta}{J})\approx \frac{\pi}{J} $ (Fig. \ref{fig1}c, green dashed). 
For small $|J-\Delta|$ the period $\tilde{t}_p$ diverges as
$\tilde{t}_p\approx \frac{1}{\Delta}\log \frac{\Delta}{|J-\Delta|}\to \infty$ signaling  a critical regime at 
 $J=\Delta$. At this point  the contrast $C(\tilde t)=\text{sech}(\Delta \tilde{t})$ monotonically decreases 
on the short timescale $\tilde{t}_{i}=\frac{\sqrt{2}}{\Delta}$ (Fig. \ref{fig1}b, green dashed).
 
To complete the characterization of the classical spin dynamics, the trajectory of the single spin direction 
$\vec{n}_1(\tilde t)$ is plotted in Fig.~\ref{fig3} (full red line) for each regime.
For later comparison with quantum results, we quote the analytical asymptotic results:
$\vec{n}_1(\tilde t)\approx[\cos \Delta \tilde{t},
 \sin\Delta \tilde{t} ,0]$ when $\frac{J}{\Delta}\ll 1$, 
and $\vec{n}_1(\tilde t)\approx[1-\frac{\Delta^2}{2J^2}\sin^2 (J\tilde{t}), \frac{\Delta}{J}\sin(J\tilde{t})\cos(J\tilde{t}),\frac{\Delta}{J}\sin^2 (J\tilde{t})]$
 when $\frac{J}{\Delta}\gg 1$.

Below, when discussing quantum dynamics, we shall still refer to the three regimes as dephased ($J\ll \Delta$), 
critical ($J\sim \Delta$) and synchronized ($J\gg \Delta$), even if synchronization {\it stricto sensu} does not occur.

\section{Quantum dynamics}
We now consider the quantum dynamics. In order to conveniently compare with the classical results, we also define the
rescaled quantities $\tilde t$, $J$ and $\Delta$, however now $\Lambda_s=\sqrt{S(S+1)}$. \footnote{The difference 
with the classical rescaling reflects the fact that the norm of a quantum spin $\sqrt{\vec{S}_j^2}=S\sqrt{(1+\frac{1}{S})}$ is not exactly $S$.} 
We emphasize that after rescaling the quantum model still explicitly depends on the spin size $S$ 
since the latter fixes the size of the Hilbert space as $(2S+1)^2$. 
In contrast to the classical case, the quantum dynamics is controled by two dimensionless parameters: $J/\Delta$ and $S$. 
In the following, we study the quantum dynamics of the three quantities of interest $C(\tilde t),C_1(\tilde t),\vec{n}_1(\tilde t)$ 
using five different and complementary approaches.

Before going to the quantitative results, we first  sketch the typical behavior of the quantum contrast $C(\tilde t)$ as a function of time, as illustrated in  Fig.~\ref{fig1}a,c (``typical'' meaning ``away from the critical case $J/\Delta$''). 
At short time, independently of $J$ and $S$, quantum and classical contrast coincide that is to say  $C(\tilde t)\approx 1- \tilde{t}^2/\tilde{t}_i$ 
with $\tilde{t}_i=\frac{\sqrt{2}}{\Delta}$. We will therefore barely discuss this short time regime from now on. 
On longer time scale, the quantum contrast $C(\tilde t)$ exhibits fast oscillations that are modulated by an envelope. 
As we are going to show, the fast oscillations essentially encode the dynamics  of the single spin direction $\vec{n}_1(\tilde t)$, or  more properly the dynamics of $n_1 ^x(\tilde t)$. To leading order, this dynamics strongly ressembles the classical dynamics and is characterized by an oscillation time scale $\tilde{t}_o$ of the order of the classical period $\tilde{t}_p$ (see Fig. \ref{fig1}a,c inset).  
By contrast, the envelope encodes the quantum dynamics of the single spin contrast $C_1(\tilde t)$ that has no classical counterpart. For large value of $S$, this envelope exhibits a very rich multi-scales dynamics with many $S$ dependent quantum time scales. 
In this work, as indicated by the arrows in Fig.~\ref{fig2}a,c, we concentrate on two of these quantum time scales: the envelope time $\tilde{t}_e$ 
where the envelope first vanishes and the approximate reccurence time $\tilde{t}_{ar}$ where it almost recovers its maximal value.

\section{Quantum spins $1/2$}
The case of two spins $S=\frac{1}{2}$ can be solved exactly as the eigenvalues and eigenvectors are easily obtained analytically \cite{long}. The contrast is 
\be
C(\tilde t)=\left| \cos(\frac{J\tilde{t}}{\sqrt{3}})\cos(\frac{\omega\tilde{t}}{\sqrt{3}} )+\frac{J}{\omega}\sin(\frac{J\tilde{t}}{\sqrt{3}})\sin(\frac{\omega\tilde{t}}{\sqrt{3}})\right|
\label{chalf}
\ee
where $\omega\equiv \sqrt{J^2+3\Delta^2}$. 
The single spin quantities $C_1(\tilde t)$ and $\vec{n}_1(\tilde t)$ can also be computed analytically \cite{long}.
The product $C_1(t)\vec{n}_1(\tilde t)=\frac{\langle \vec{S}_{1} \rangle}{S}$ is plotted in Fig. \ref{fig3} and will be discussed in a later part. 
The contrast $C(\tilde t)$ given by Eq. (\ref{chalf}) exhibits
 fast oscillations modulated by an envelope. Because of the latter, the contrast always vanishes at some time whatever the ratio $J/\Delta$. 
This means that there is no synchronization transition in this $S=\frac{1}{2}$ quantum case. 
However, there are still two distinct regimes separated by a crossover around $J=\Delta$:\\ 
(i) For $J\ll \Delta$, the contrast is
\be
C(\tilde t)\approx \left|\cos(\Delta \tilde{t})\cos(\frac{J\tilde{t}}{\sqrt{3}})\right|.
\label{chalfsmallj}
\ee
It features fast oscillations leading to a periodic vanishing of the contrast with a period $\tilde{t}_o=\frac{\pi}{\Delta}$. These fast oscillations 
correspond to a single spin direction identical to the classical result $\vec{n}_1(\tilde t)\approx[\cos(\Delta \tilde{t}),\sin(\Delta \tilde{t}),0]$. 
The contrast envelope is given by the single spin contrast $C_1(\tilde t)=|\cos(\frac{J\tilde{t}}{\sqrt{3}})|$
and is characterized by an envelope time $\tilde{t}_e=\frac{\pi \sqrt{3}}{2J}$ and a reccurence time $\tilde{t}_{ar} \sim 2\tilde{t}_e$. 
For a generic value of $J/\Delta$ the relevant Bohr frequencies are incommensurate and the contrast is only quasi-periodic.\\
(ii) For $J\gg \Delta$, the contrast is
\be
C(\tilde t)\approx \left|\cos(\frac{\sqrt{3}\Delta^2}{2J}\tilde{t})[1-\frac{3 \Delta^2}{2J^2}\sin^2(\frac{J\tilde{t}}{\sqrt{3}})]\right| .
\label{chalflargej}
\ee
The fast oscillations of period $\tilde{t}_o=\frac{\pi\sqrt{3}}{J}$, have now a small amplitude of order $\Delta^2/J^2$ that do not lead to a vanishing contrast. 
These  oscillations are well in correspondance with the component $|n_1^x(\tilde t)|$ of
the single spin direction $\vec{n}_1(\tilde t)=\textrm{sign}(\cos(\frac{\sqrt{3}\Delta^2}{2J}\tilde{t}))[1-\frac{3 \Delta^2}{2J^2}\sin^2(\frac{J\tilde{t}}{\sqrt{3}}),
\frac{\sqrt{3} \delta}{2J}\sin(\frac{J\tilde{t}}{\sqrt{3}})\cos(\frac{J\tilde{t}}{\sqrt{3}}),\frac{\sqrt{3} \delta}{2J}\sin^2(\frac{J\tilde{t}}{\sqrt{3}})]$ 
(the latter sharing the same structure as the classical result $\vec{n}_1$ except for the global sign function). 
The contrast envelope corresponds to a single spin contrast $C_1(\tilde t)=|\cos(\frac{\sqrt{3}\Delta^2}{2J}\tilde{t})|$ with an envelope time 
 $\tilde{t}_e=\frac{\pi J}{\sqrt{3}\Delta^2}$ and a reccurence time $\tilde{t}_{ar} \sim 2\tilde{t}_e$. The contrast still vanishes but on the much longer 
timescale $\tilde{t}_e$, therefore mimicking synchronization during the short time dynamics. The timescale $\tilde{t}_e \sim \frac{J}{\Delta^2}$ is reminiscent of off-resonance Rabi flopping \cite{gibble}. \\
(iii) The crossover from one regime to the other occurs in the vicinity of 
$J=\Delta$ where the contrast has a simple expression 
$C(\tilde t)=|\cos(\frac{\Delta}{\sqrt{3}} \tilde{t})|^3$ such that all time scales are of the same order $\tilde{t}_o\sim \tilde{t}_e\sim \frac{1}{\Delta}$. 

\section{Numerics}
In order to study the influence of the spin size, for $S$ between $1$ and $20$ and various $J/\Delta$ values, we  numerically obtain the eigenvalues and eigenvectors of the time-independent Hamiltonian of size $(2S+1)^2\times (2S+1)^2$ and 
then compute the single spin  time evolution $\langle \vec{S}_{1}(\tilde t)\rangle$, that contains all the necessary information. 
As illustrated in Figs.~\ref{fig1}a,c, for $S=10$, we obtain two regimes of behavior for the contrast $C(\tilde t)$ depending on $J$ being larger or smaller than $\Delta$. 
For $J \ll \Delta$, the contrast has fast oscillations leading to a periodically vanishing contrast with the oscillation period $\tilde{t}_o \sim 1/\Delta$ 
(Fig. \ref{fig1}a). These oscillations are modulated by an envelope with an envelope time that can be fitted as $\tilde{t}_{e} \sim \sqrt{S}/J$.
For $J \gg \Delta$, the contrast features fast oscillations of small amplitude and with oscillation period $\tilde{t}_o \sim 1/J$ (Fig. \ref{fig1}c). 
In this regime the contrast  only vanishes with the envelope on a much longer time scale $\tilde{t}_e\sim \sqrt{S} J/\Delta^2$. 
Extrapolating these two asymptotic behaviors, the ratio $\tilde{t}_{e}/\tilde{t}_{o}$ is minimal for $J\sim \Delta$ 
($\tilde{t}_{o}$ is maximal and $\tilde{t}_{e}$ is minimal). In fact, in this crossover regime
$J\sim \Delta$ and for the studied values of $S$, it appears difficult to correctly distinguish the two time scales $\tilde{t}_o$ and $\tilde{t}_{e}$. 
As illustrated in Figs.~\ref{fig1}b, after an initial short time decay, the contrast fluctuates a lot and we mainly observe
that the average amplitude of these fluctuations decreases with increasing $S$.

\section{Large $S$ approach}
To study whether quantum fluctuations destroy the classical synchronization for any finite $S$ (even arbitrary large), 
we have developed a large $S$ cumulant approach following \cite{Garanin}. We refer to \cite{long} for the presentation of these lengthy calculations. Here, we just plot the contrast obtained using the cumulant approach and compare it with the classical dynamics and the quantum numerics in Fig.~\ref{fig1} (full red line). We observe that the cumulant approach deviates from the classical dynamics and correctly follows the quantum dynamics up to a critical time $\tilde{t}_c$ at which it deviates from the latter by overestimating quantum corrections. In the dephased regime $J\ll \Delta$ we find $\tilde{t}_c \sim \tilde{t}_e$ whereas in the synchronized regime  $J\gg \Delta$, $\tilde{t}_c \sim \tilde{t}_e/20$ .

\section{Effective Hamiltonian at small $J_s/\Delta_s$}
\begin{figure}[ht]
\onefigure[width=6cm]{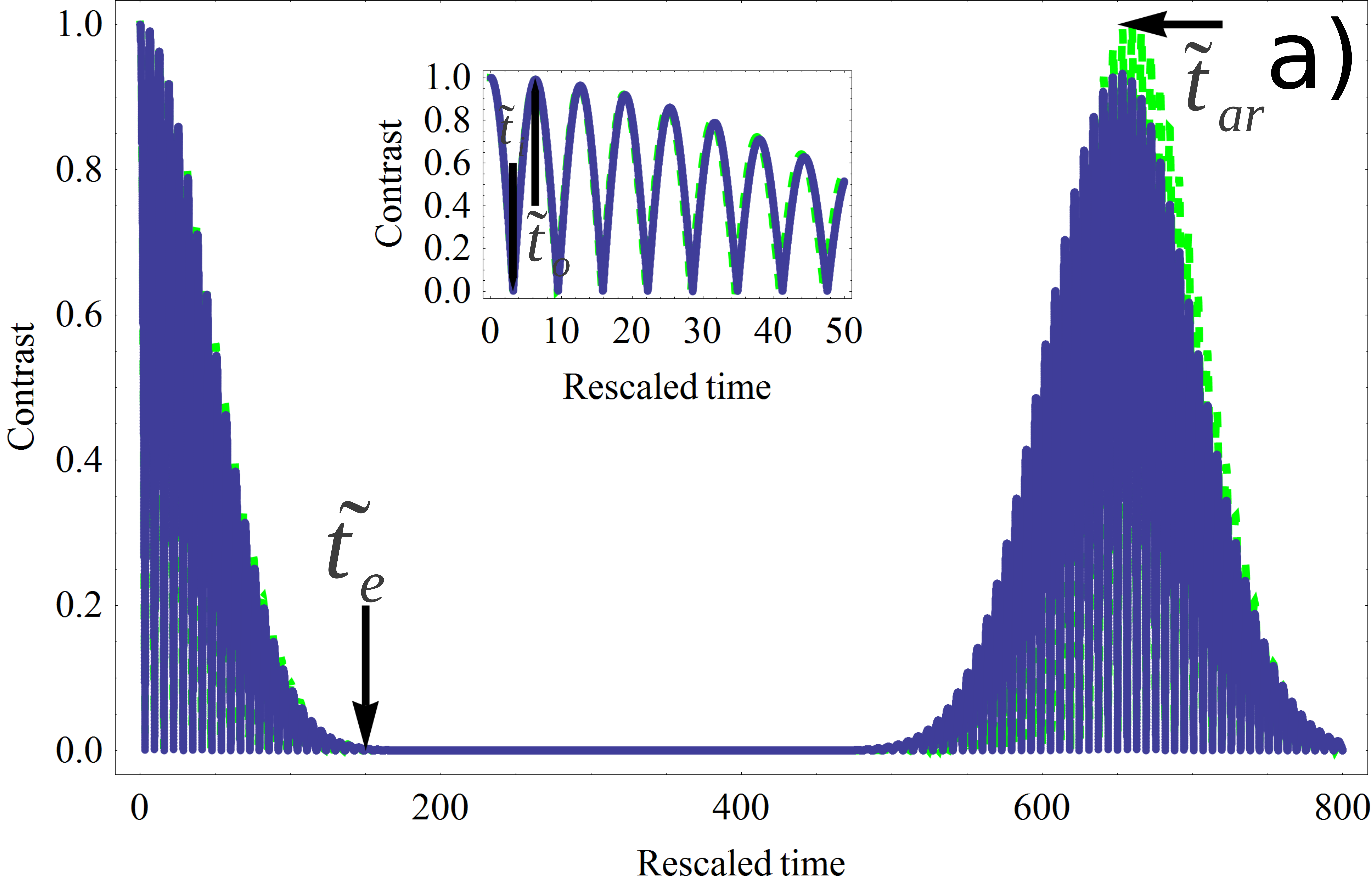}
\onefigure[width=6cm]{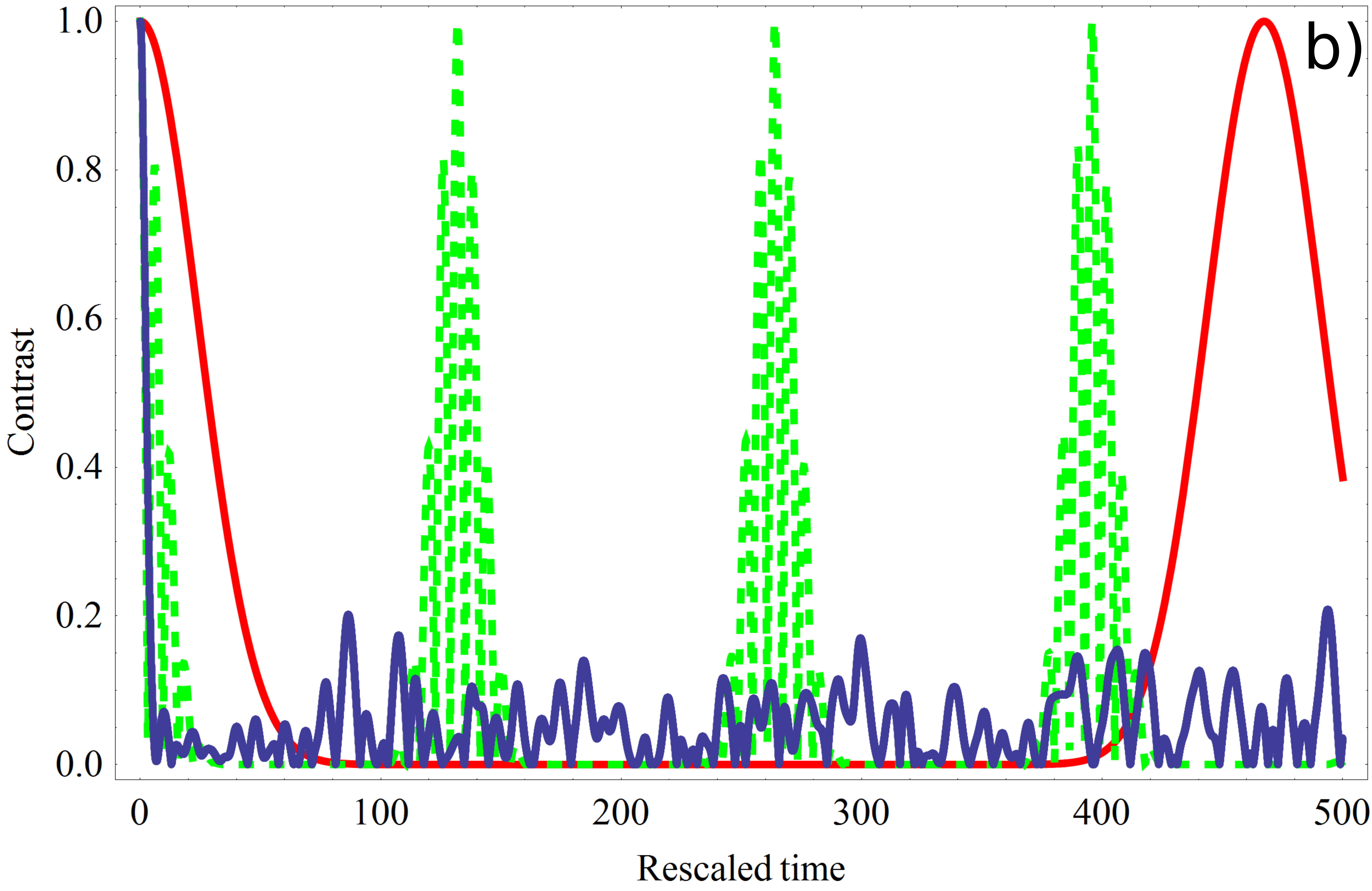}
\onefigure[width=6cm]{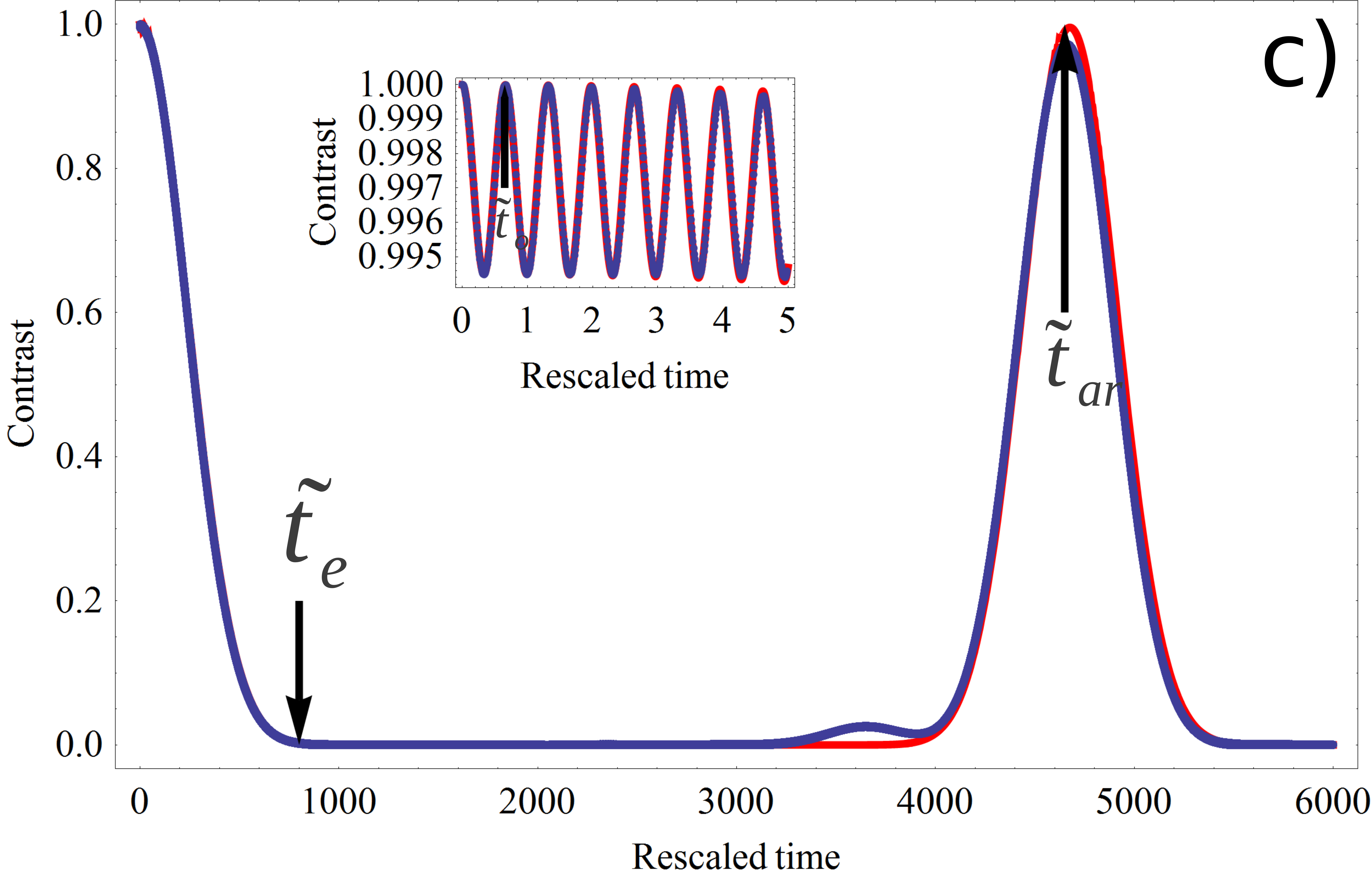}
 \caption{Contrast $C$ as a function of rescaled time $\tilde{t}$ [units of $\frac{1}{\Delta}$] for $S=10$. Comparison between small $J_s/\Delta_s$ (dashed green line), 
large $J_s/\Delta_s$ (full red line) approaches and quantum numerics (dotted blue line) in 3 regimes: (a) $J/\Delta=0.2$ (dephased); (b) $J/\Delta=1$ (critical) and (c) $J/\Delta=10$ 
(synchronized). Initial decay ($\tilde{t}_i$), fast oscillations ($\tilde{t}_o$), envelope ($\tilde{t}_e$) and approximate recurrence ($\tilde{t}_{ar}$) times are indicated.}
\label{fig2}
\end{figure}
In the limit $J_s/\Delta_s\ll 1$, the dynamics in the complete Hilbert space appears to be well described by the effective Hamiltonian $H_{J_s\ll \Delta_s}=\Delta_s (S_1^z-S_2^z)+J_s S_1^z S_2^z$ \cite{long}. Its eigenvectors and eigenvalues are $|S_1=S,m_1;S_2=S,m_2\rangle$ and $E_{m_1,m_2}=\Delta_s (m_1-m_2)+J_s m_1 m_2$ with $m_j=-S,\ldots,S$. In that situation we find that the contrast is 
\be
C(\tilde{t})=\left|\cos(\Delta \tilde{t})\cos^{2S}(\frac{J \tilde{t}}{2\sqrt{S(S+1)}})\right|,
\label{csmallj}
\ee
which recovers Eq. (\ref{chalfsmallj}) when $S=1/2$. Equation (\ref{csmallj}) 
features fast oscillations leading to a periodic vanishing of the contrast with a period $\tilde{t}_o=\frac{\pi}{\Delta}$. 
These oscillations correspond to a single spin direction identical to the classical and spin $1/2$ results: $\vec{n}_1(\tilde t)\approx[\cos(\Delta \tilde{t}),\sin(\Delta \tilde{t}),0]$.
Eq.~(\ref{csmallj}) reveals an envelope corresponding to a single spin contrast $C_1(\tilde{t}) \approx \left|\cos^{2S}(\frac{J \tilde{t}}{2\sqrt{S(S+1)}})\right|$ 
such that for large but finite $S \gg 1$ it predicts an envelope time $\tilde{t}_e\approx \frac{2\sqrt{S}}{J}$ 
and an envelope reccurence time $\tilde{t}_{ar}=\frac{2\pi \sqrt{S(S+1)}}{J}\approx \pi \sqrt{S} \tilde{t}_e$. 
Comparison to numerics is excellent for times $\tilde{t} \lesssim \tilde{t}_{ar}$, see Fig. \ref{fig2}a. 
Numerics show however that the reccurence is only approximate, implying the existence of longer characteristic quantum time scales not accessible within this analytical approach.
Note that when $S\to \infty$, then  $\tilde{t}_e \to \infty$ and the classical result $C_1(\tilde t)=1$ is recovered.

\section{Effective Hamiltonian at large $J_s/\Delta_s$}
In the limit $J_s/\Delta_s\gg 1$, using the Bloch-Horowitz projection method \cite{BH}, we obtain an effective Hamiltonian 
$H_{J_s\gg \Delta_s}=\frac{J_s}{2}[ \vec{S}_t^2-2S(S+1)\mathbb{I}] + \frac{\Delta_s^2}{J_s 2S(4S-1)}[(2S)^2\mathbb{I}-(S_t^z)^2]$ where $\vec{S}_t=\vec{S}_1+\vec{S}_2$.
It describes the dynamics in a truncated Hilbert space restricted to the largest Bloch sphere of radius $2S$ 
\cite{long}. 
Its eigenvectors and eigenvalues are $|S_t=2S;M\rangle$ and $E_M=- \frac{\Delta_s^2}{J_s 2S(4S-1)}M^2+\textrm{ct}$ with $M=-2S,\ldots,2S$. 
This approach automatically integrates over the fast oscillations and only captures the envelope giving a contrast 
$C(\tilde{t})= C_1(\tilde{t})= \left| \cos^{4S-1}(\frac{\Delta^2 \tilde{t} \sqrt{S(S+1)}}{2J S(4S-1)})\right|$. 
Using a slightly different approach, which consists in keeping the two largest Bloch sphere ($2S$ and $2S-1$) in a truncated Hilbert space \cite{long}, we find that the contrast is actually
\begin{eqnarray}
C(\tilde{t})\approx \left| \cos^{4S-1}(\frac{\Delta^2 \tilde{t} \sqrt{1+\frac{1}{S}}}{2J (4S-1)})\right. \times \nonumber \\
\left. [1-\frac{\Delta^2(1+\frac{1}{S})}{2 J^2}\sin^2(\frac{J\tilde{t}}{\sqrt{1+\frac{1}{S}}})]\right|,
\label{clargej} 
\end{eqnarray}
which recovers Eq. (\ref{chalflargej}) when $S=1/2$.
Equation (\ref{clargej}) features fast oscillations with period ${\tilde t}_o=\frac{\pi \sqrt{1+1/S}}{J}$ and small amplitude of order $\frac{\Delta^2(1+\frac{1}{S})}{2 J^2}$ 
that coincide with the classical result up to $1/S$ corrections.
For large but finite $S \gg 1$, the single spin contrast $C_1(\tilde{t})$ predicts an envelope time $\tilde{t}_e\approx 4\frac{\sqrt{2S}J}{\Delta^2}$
and an envelope reccurence time $\tilde{t}_{ar}=\frac{2\pi J S(4S-1)}{\Delta^2\sqrt{S(S+1)}}\approx\pi  \sqrt{2S} \tilde{t}_e$.
As in the previous case, the agreement with numerics is very good for times $\tilde{t} \lesssim \tilde{t}_{ar}$, see Fig. \ref{fig2}c. The main result of this section is that the contrast first vanishes at the envelope time $\tilde{t}_e \sim \sqrt{S}J/\Delta^2$, similar to the $S=1/2$ case but that diverges when $S \to \infty$, therefore pointing toward synchronization in the classical limit. The physical mechanism behind $\tilde{t}_e \sim \sqrt{S}J/\Delta^2$ is that of virtual transitions to Bloch spheres of smaller radius leading to dephasing.
Nevertheless, for large but finite $S \gg 1$, the quantum aspect is striking when considering the norm of the total spin: 
on the one hand $\sqrt{\langle\vec{S}_t^2\rangle}=\sqrt{2S(2S+1)}$ is a constant, 
as the system remains on the largest Bloch sphere, but on the other $|\langle \vec{S}_t \rangle|$ goes to zero at the envelope time. 
These quantities correspond to two definitions of the norm of a vector which coincide classically but not quantum mechanically. 

As illustrated in Fig.~\ref{fig2}b, we note that neither Eq. (\ref{csmallj}) nor Eq. (\ref{clargej}) correctly captures the quantum contrast around the critical point $J/\Delta=1$ of the classical dynamics.

\section{Single spin on the Bloch sphere}
\begin{figure}[ht]
\includegraphics[width=2.8cm]{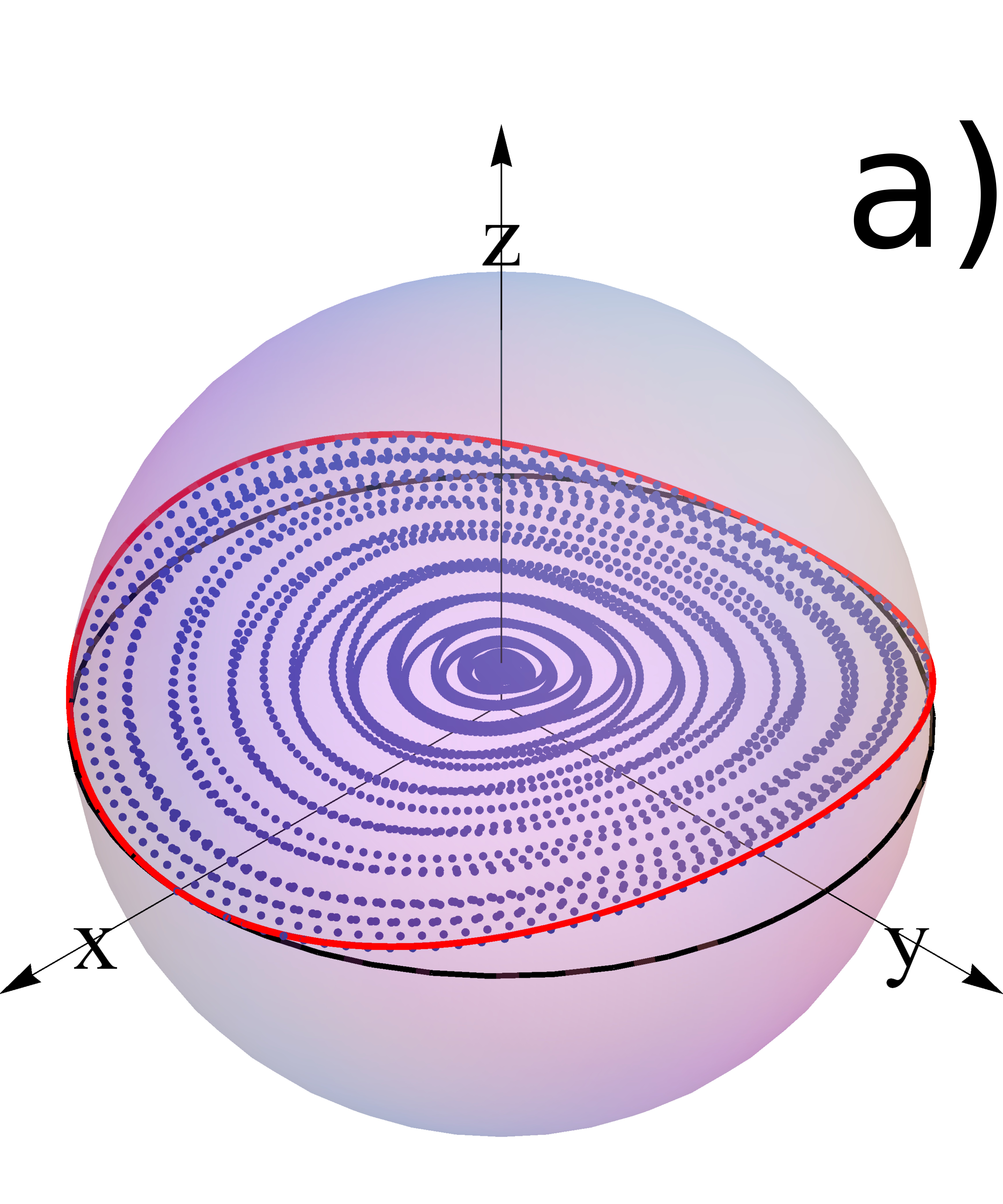}
\includegraphics[width=2.8cm]{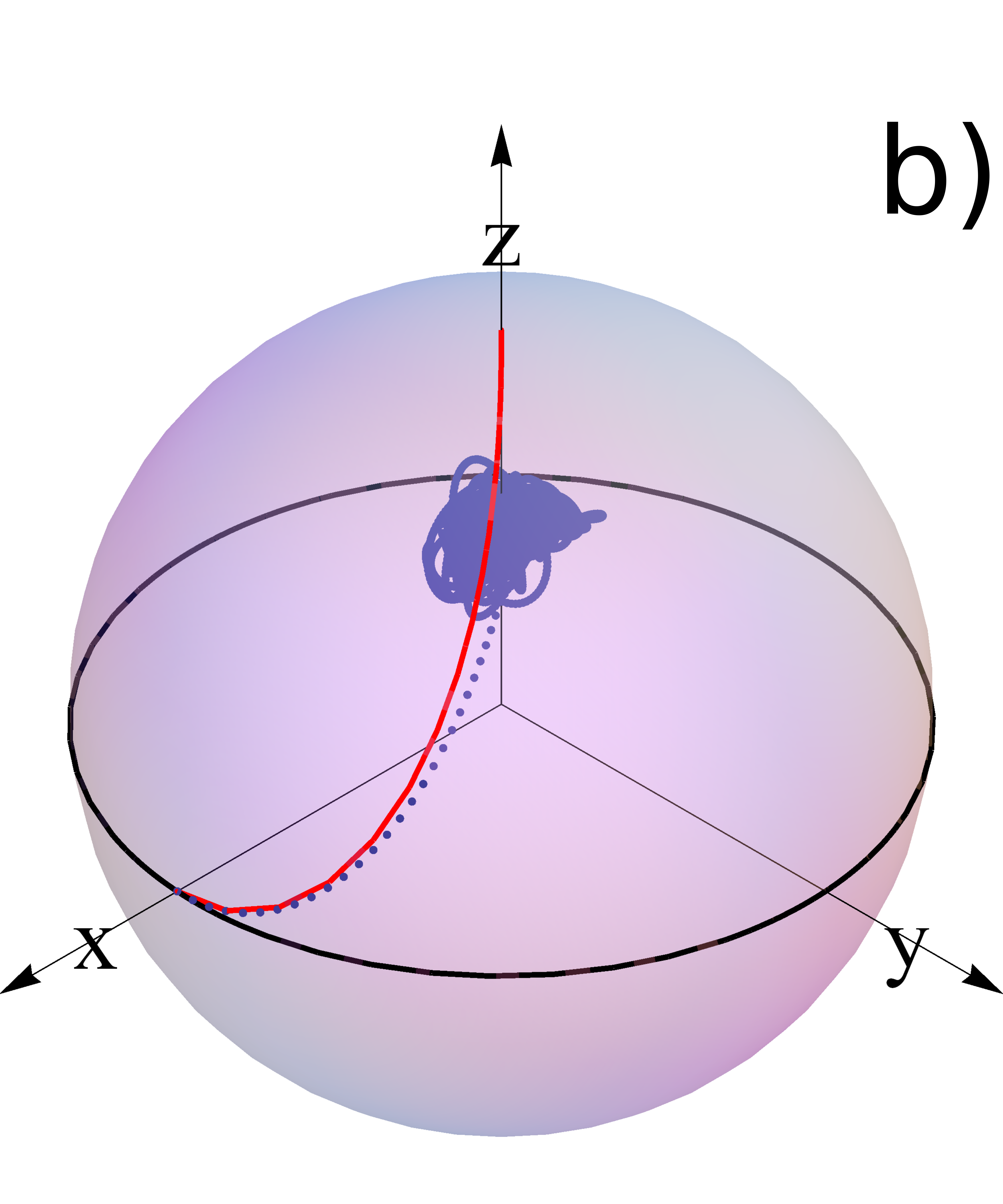}
\includegraphics[width=2.8cm]{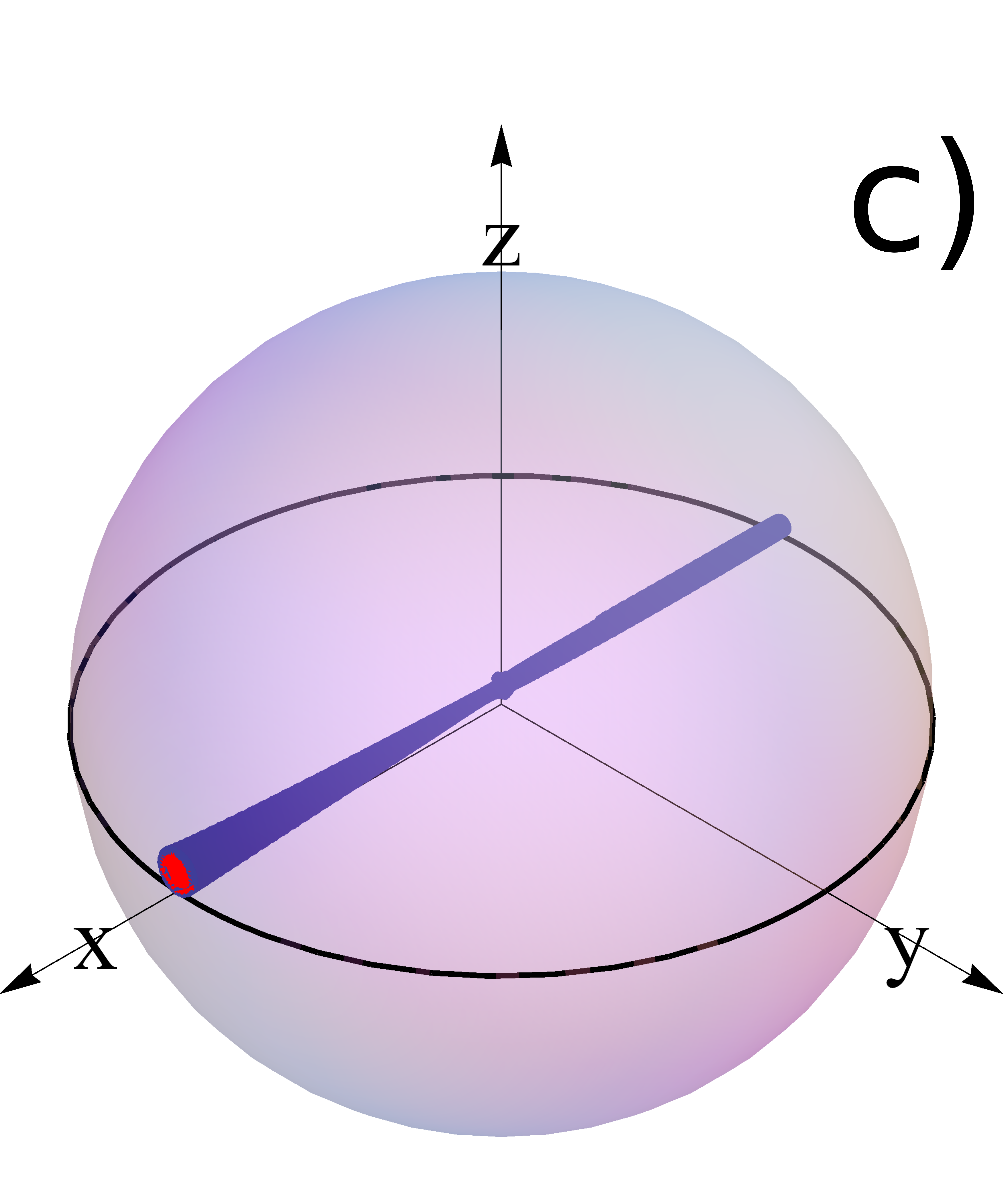}\\
\includegraphics[width=2.8cm]{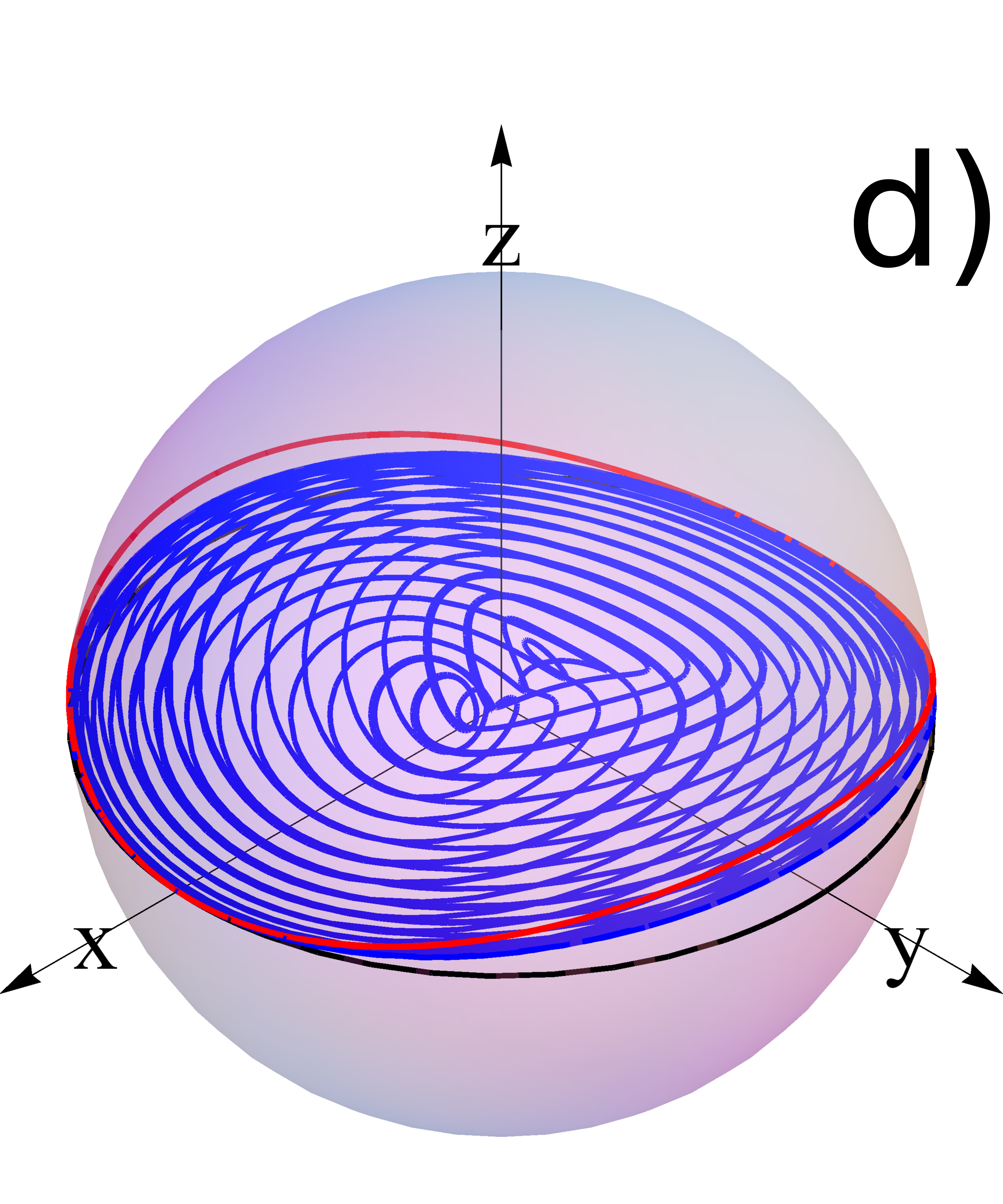}
\includegraphics[width=2.8cm]{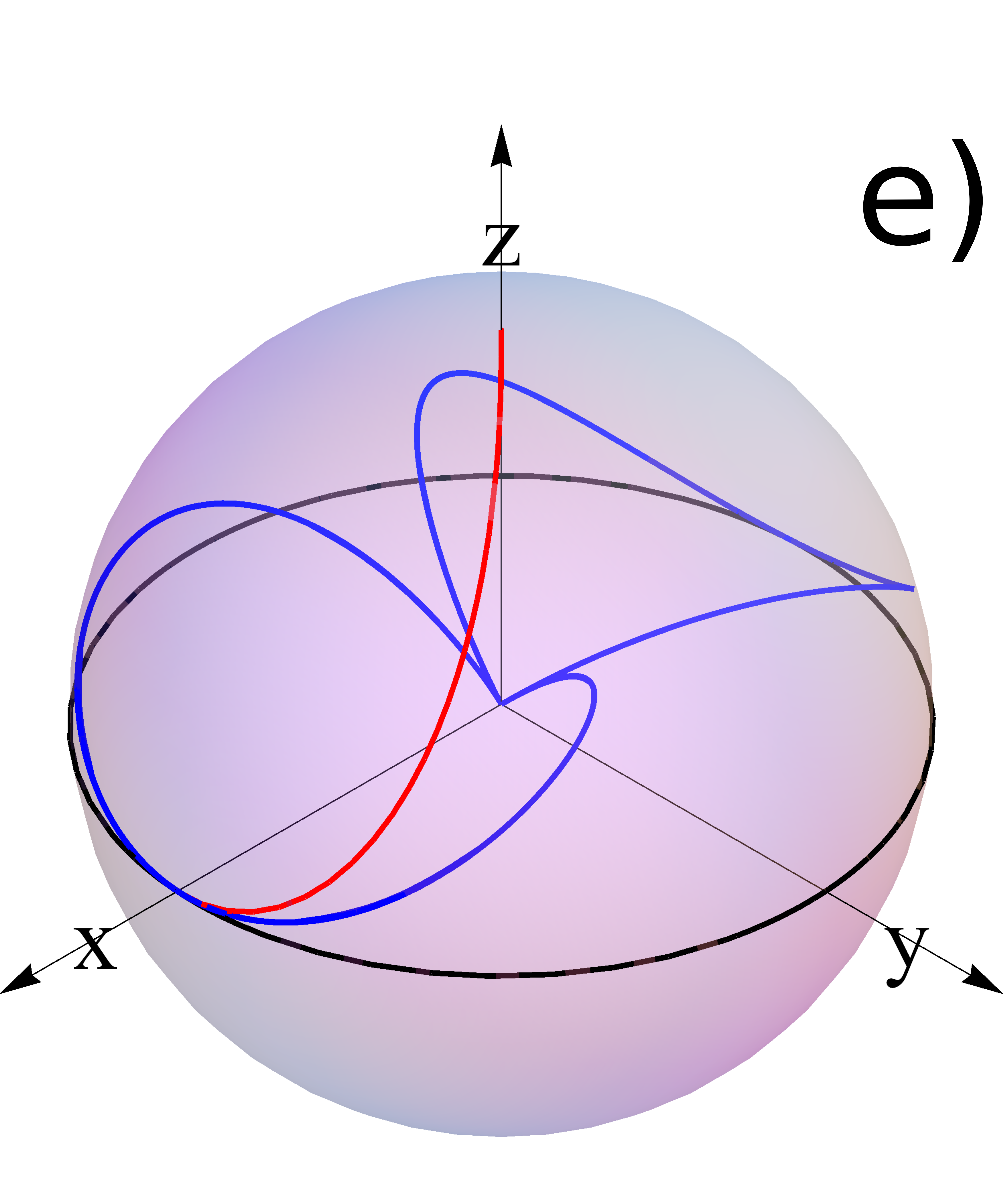}
\includegraphics[width=2.8cm]{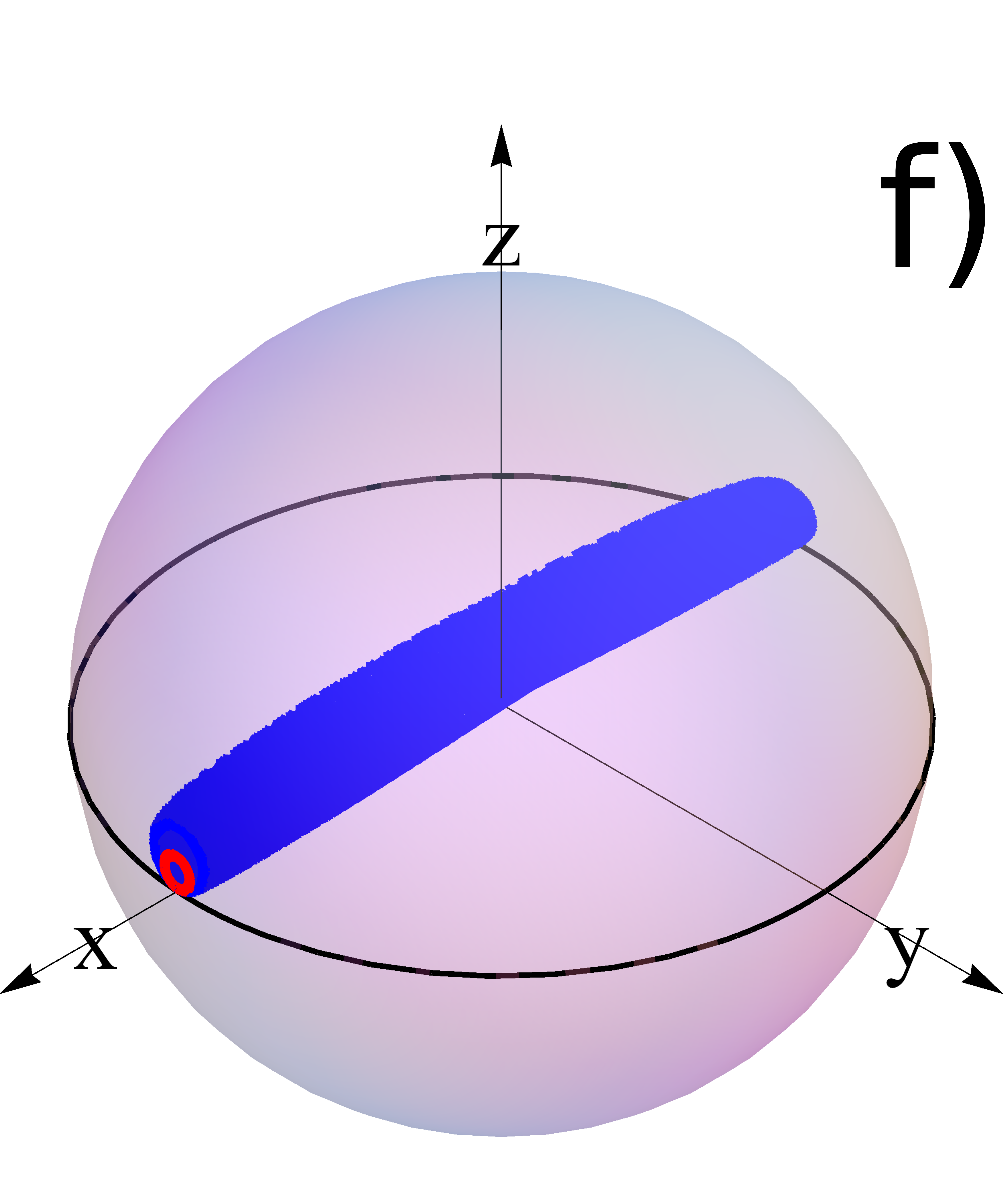}
 \caption{Dynamics of a single spin $\langle \vec{S}_1 \rangle/S=C_1 (\tilde t)\vec{n}_1(\tilde t)$ on the Bloch sphere 
for times $0\leq \tilde{t}\leq \tilde{t}_{ar}$. Comparison between quantum (dotted blue line) and classical (full red line) 
dynamics in 3 regimes: First column (a,d): $J/\Delta=0.2$ (dephased); second column (b,e): $J/\Delta=1$ (critical); 
third column (c,f): $J/\Delta = 10$ (synchronized). First line (a,b,c): $S=10$; second line (d,e,f): $S=1/2$. The equator is indicated with a black line.}
\label{fig3}
\end{figure}
The single spin representation $\langle \vec{S}_1\rangle/S = C_1 (\tilde t)\vec{n}_1(\tilde t)$ is particularly powerful in providing a physical picture 
for the system's dynamics and in comparing the classical and the quantum cases. Figure \ref{fig3} show the single spin behavior 
on the Bloch sphere in three regimes -- dephased $J\ll \Delta$, critical $J=\Delta$ and synchronized $J\gg \Delta$ -- 
for both the classical (red lines) and  quantum dynamics (blue dots) and for two spin sizes $S=10$ and $S=1/2$. 
Away from $J/\Delta=1$, the dynamics of the single spin direction $\vec{n}_1(\tilde t)$ is quite similar in the classical and quantum cases: 
in the dephased regime, it stays almost in the equatorial plane 
and turns around $z$ axis with a period $\tilde t_{o}\approx \pi/\Delta$ (see Fig. \ref{fig3}a,d); in the synchronized regime, apart from small ($S$ dependent) amplitude oscillations of period $\tilde t_{o}\approx \pi/J$, the direction $\vec{n}_1(\tilde t)$ is almost locked and aligned with $+x$ except 
that in the quantum case it jumps back and forth from $+x$ to $-x$ at every odd multiple of the approximate reccurence time $\tilde{t}_{ar}$ (see Fig.~\ref{fig3}c,e and the sign function in $\vec{n}_1$ below Eq. (\ref{chalflargej})). 
Most of the quantum behavior is thus encoded in the dynamics of the single spin contrast $C_1(\tilde t)$ 
which  vanishes at the envelope time $\tilde{t}_e$ and almost recovers at the reccurence time $\tilde{t}_{ar}$, 
where the $S$ dependent timescales $\tilde{t}_e$ and $\tilde{t}_{ar}$ depend on the considered regime. 
In the dephased regime the combined dynamics of $C_1(\tilde t)$ and $\vec{n}_1(\tilde t)$ results 
in an inwards spiral motion for the single spin in the equatorial plane (see Fig. \ref{fig3}a,d) until $\tilde{t}_e$, 
followed by an outwards spiraling until $\tilde{t}_{ar}$. For $S=10$, the long time interval $\tilde{t}_e<\tilde t<\tilde{t}_{ar}$ during which $C_1(\tilde t)\simeq 0$ corresponds to the higher density of points visible in the center of the Bloch sphere in Fig.~\ref{fig3}a. 
In the synchronized regime, the combined dynamics now results in a cigar-like shape of the complete trajectory visible on Fig.~\ref{fig3}c,f. In the critical case, the classical direction $\vec{n}_1$ goes monotonically from $x$ to $z$.
The quantum dynamics  is here quite different and depends strongly on $S$. More precisely, 
when increasing $S$, it evolves towards the classical one for times shorter than the envelope time (see Fig. \ref{fig3}b,e), however at longer times, the quantum single spin always ends up visiting the inner part of the Bloch sphere, while the classical spin is constrained to its surface.

\section{Qualitative picture}
The main qualitative picture emerging from our study is that the dynamics of a quantum spin (and of the quantum constrast) results from the combined dynamics of its effective direction $\vec{n}_1(\tilde t)$ and its effective norm or envelope $C_1(\tilde t)$. On the one hand, the short and fast time dynamics are governed by the direction $\vec{n}_1(\tilde t)$ and, apart from $1/S$ corrections, it retains most of the classical aspect of the spin dynamics; in particular $\tilde{t}_i=\frac{\sqrt{2}}{\Delta}$ is the same in the quantum and classical cases and does not depend on $J$ or $S$. In addition, away from $J=\Delta$, the fast oscillations time $\tilde{t}_o\approx \textrm{min}(\frac{\pi}{\Delta},\frac{\pi}{J})$ corresponds well to the classical period $\tilde{t}_p$.
On the other hand, the slow and long times quantum spins dynamics are governed by the envelope $C_1(\tilde t)$. 
Since  $C_1(\tilde t)$ is a direct measure of the effective spreading of the spin wavepacket on the Bloch sphere,
this part of the dynamics is a pure quantum phenomenum and it strongly depends on $S$. 
For large value of $S$, this envelope $C_1(\tilde t)$ has a very rich multiscale dynamics featuring
collapses ($C_1\sim 0$), revivals and reccurences ($C_1\sim 1$). Quantitatively, we found that, away from  $J=\Delta$, the 
 envelope time of the first collapse $\tilde{t}_e\sim \sqrt{S}\textrm{ max}(\frac{1}{J},\frac{J}{\Delta^2})$ and the time of the first approximate recurrence
 $\tilde{t}_{ar}\sim  S\textrm{ max}(\frac{1}{J},\frac{J}{\Delta^2})\sim \sqrt{S}\tilde{t}_e$ both diverge with $S$ in the classical limit. 
More generally, we expect an increasing hierarchy of collapse and reccurence quantum timescales with increasing values of $S$.

The way the classical limit emerges from the quantum dynamics as $S\to \infty$ is thus quite remarkable. The classical dynamics is characterized by only two timescales ($\tilde{t}_i$ and $\tilde{t}_p\sim \tilde{t}_o$), whereas the quantum dynamics has more and more timescales as $S$ increases ($\tilde{t}_e$, $\tilde{t}_{ar}$, etc.), so that a priori it seems unlikely that the two pictures would reconcile in the large $S$ limit. 
However all of the quantum timescales diverge when $S\to \infty$ so that in the end, the classical dynamics emerges as the short time behavior of the quantum dynamics.
 The transition occurs between the oscillation $\tilde{t}_0$ and the envelope $\tilde{t}_e$ times. The situation close to the classical critical point $J/\Delta=1$ 
is quite favorable to observe quantum effects, as the classical period diverges (see Fig. \ref{fig2}b) and the envelope time is minimal, although it is difficult to study analytically.

A similar behavior -- namely multi-scale quantum dynamics and phase transition in the classical dynamics only -- was found in a simpler non-linear {\it single} large spin model. This is the Bose-Hubbard dimer, describing e.g. a Bose-Einstein condensate in a double-well potential \cite{Milburn}, and which, in the spin language, maps onto the Lipkin-Meshkov-Glick model, see e.g. \cite{vidal}. In the classical or mean-field limit the condensate is described by a Gross-Pitaevskii equation, 
which features a self-trapping transition \cite{Eilbeck}. At small onsite repulsion, the bosons initially all in the same well can tunnel to the other well.
 But when the onsite repulsion exceeds a critical value, the bosons are essentially trapped in their initial well. Including quantum fluctuations has a dramatic effect:
 the self-trapping transition no more exists -- for all interaction strength, the condensate can visit both wells -- 
and a rich multi-scale quantum dynamics emerges, see e.g. \cite{Kalosakas}. The complete dynamics involves at least three very different timescales, 
the two largest being quantum and increasing very fast with the spin size \cite{Kalosakas}. 

\section{Experimental relevance}
In \cite{deutsch}, a spin kinetic equation in energy space $\vec{S}(E,t)$ was developed and solved numerically in order to compare to the experimental results. The agreement between theory and experiment was quite good but the spins were intrinsically classical. Here, the two spins model captures the physics of the synchronization transition in the large $S$ limit and also includes the effect of quantum fluctuations. In the dephased regime, it features spurious large amplitude fast oscillations, which are due to the huge simplification in the size of the Hilbert space from the microscopic dimension $2^N$ to the effective dimension $(2S+1)^2=(N/2+1)^2$. The most interesting aspect of the two spins model is the appearance of quantum dynamics on the timescale of the envelope $\tilde{t}_e$. From \cite{deutsch}, we get $\Delta \sim 10^5$ rad/s, $J/\Delta\sim 1-10$ (synchronized regime) and $S=N/4\sim 10^4$ so that the envelope time $t_e\approx \tilde{t}_e S\sim 10-100$~s is larger but comparable to the duration of the experiment ($\sim 5$ s). It would be interesting to measure longer timescales to detect the appearance of quantum dynamics modulating the classical behavior by sequences of collapses and revivals. The experiment of \cite{deutsch} may be even further in the synchronized regime with $J/\Delta$ up to $\sim 100$. For another experiment with cold atoms \cite{thomas}, we find $J/\Delta \sim 0.25$ (dephased regime). The two spins model is not only relevant to cold atoms \cite{deutsch,kleine,thomas}, but should also concern other experiments such as two quantum dots trapping spins \cite{burkard}, two nanodisks with large magnetic moments \cite{Pigeau} 
or generally systems with two coupled macrospins such as molecular magnets, in particular so-called molecular multidot devices \cite{molecularmagnets}.

\section{Conclusion and perspectives}
In conclusion, we introduced and studied a quantum model of two exchange-coupled spins $S$ in the presence of an inhomogeneous magnetic field. Our main result is that quantum fluctuations prevent the occurrence of the synchronization transition found in the classical limit. The quantum dynamics is very rich, featuring collapses, revivals and recurrences described by multiple timescales. The most important quantum feature is the envelope of the contrast, which corresponds to the spreading of a single spin wavepacket. The classical dynamics is reached in a non-trivial way emerging as the short time behavior of the quantum dynamics. The quantum dynamics at the classical critical point $J=\Delta$ is especially intriguing and deserves further investigations. In the future, we plan to study the possibility of spin squeezing  \cite{reviewspinsqueezing} within our simple model. It would also be interesting to include the effect of coupling to a bath that might, among other effects, favor a synchronization transition by rendering the system more classical \cite{lehur}.


\acknowledgments
We thank G. Ferrini, J. Vidal, B. Dou\c{c}ot, K. Le Hur, F. Lalo\"e, P. Rosenbusch and J. Reichel for useful discussions.

\end{document}